\documentstyle[12pt]{article}
\newcommand{\Frac}[2]%
{{\textstyle \frac{\mbox{\footnotesize $#1$}\rule[-0.9mm]{0mm}{1mm}}%
{\mbox{\footnotesize $#2$}\rule{0mm}{3.1mm}}}}
\begin{document}
\begin{titlepage}
\vspace*{-12 mm}
\noindent
\begin{flushright}
\begin{tabular}{l@{}}
hep-ph/9711423 \\
\end{tabular}
\end{flushright}
\vskip 12 mm
\begin{center}
{\LARGE\bf
The Spin and Flavour Dependence of } \\
\vspace{3ex}
{\LARGE\bf
  High-Energy Photoabsorption}
\\[14 mm]
{\bf S.D. Bass}$^{\ a,}$\footnote{sbass@pythia.itkp.uni-bonn.de}
and {\bf M.M. Brisudov\'{a}}$^{\ b,}$\footnote{brisuda@t5.lanl.gov}
\\[10mm]   
$^a$ {\em Institut f\"{u}r Theoretische Kernphysik, 
Universit\"{a}t Bonn,\\
Nussallee 14--16, D-53115 Bonn, Germany}\\[5mm]
$^b$ {\em Theoretical Division, 
Los Alamos National Laboratory, \\ 
Los Alamos, NM 87545, U.S.A.  }
\end{center}
\vskip 10 mm
\begin{abstract}
\noindent
We review the present data on high-energy, spin-dependent
photoabsorption.
We find a strong isotriplet term in 
$(\sigma_A - \sigma_P)$ 
which persists from $Q^2 \sim 0.25$GeV$^2$ to high $Q^2$
polarised deep inelastic scattering.
For $Q^2 \sim 4$GeV$^2$ and $x$ between 0.01 and 0.12
the isotriplet part of $g_1$ behaves as
$g_1^{(p-n)} \sim x^{-{1 \over 2}}$,
in contrast to soft Regge theory which predicts that 
$g_1^{(p-n)}$ should converge as $x \rightarrow 0$.
The isotriplet, polarised structure function
$2x g_1^{(p-n)}$ 
is significantly greater than the isotriplet, unpolarised 
structure function
$F_2^{(p-n)}$ in this kinematic region.
We analyse the low $Q^2$ photoabsorption data from E-143 
and SMC
and use this data to estimate the high-energy Regge
contribution to the Drell-Hearn-Gerasimov sum-rule.

\end{abstract}
\end{titlepage}
\renewcommand{\labelenumi}{(\alph{enumi})}
\renewcommand{\labelenumii}{(\roman{enumii})}
\newpage

\section {Introduction}

Spin sum-rules for real and deeply virtual photoabsorption
provide important constraints for our understanding of the 
structure of the nucleon.
Experimental tests of these sum-rules involve some 
extrapolation of the measured cross-sections to asymptotic
$\sqrt{s}$ at fixed $Q^2$. These extrapolations are, in 
general, motivated by Regge theory or perturbative QCD.

In this paper we discuss the spin and flavour 
dependence of
high-energy photoabsorption.
Polarised deep inelastic scattering experiments at 
CERN \cite{expt,smc}, DESY \cite{hermes}
and SLAC \cite{e143a,e143b, e154} 
have measured 
the 
spin asymmetry
\begin{equation}
A_1 = {\sigma_A - \sigma_P \over \sigma_A + \sigma_P}
\end{equation}
over a large range of $x$ and $Q^2$.
Here
$\sigma_A$ and $\sigma_P$ denote the cross-sections 
for the absorption of a transversely polarised 
photon by a nucleon where the photon is polarized
anti-parallel
$\sigma_A$ and parallel $\sigma_P$ to the nucleon.
We concentrate on $A_1$ at 
small Bjorken $x$ and large $\sqrt{s_{\gamma p}}$.
We begin (Section 2) with a brief review of
the present status of
sum-rules 
for $(\sigma_A - \sigma_P)$ 
and spin dependent Regge theory.
In Section 3 we present our new results.
We examine the present data on $A_1$ 
and make several interesting observations.
\begin{enumerate}
\item
There is a strong isotriplet term in $(\sigma_A - \sigma_P)$
at large $\sqrt{s}$ which persists from 
low $Q^2$ $\sim 0.25$GeV$^2$ through to
polarised deep inelastic scattering.
\item
The isotriplet part of $g_1$
(the nucleon's first spin-dependent, 
 deep inelastic structure function)
behaves as $\sim x^{- {1 \over 2}}$ 
at small $x$ (between 0.01 and 0.12) 
and $Q^2 \sim 4$GeV$^2$.
This is in contrast to soft 
Regge theory which predicts
that
$g_1^{(p-n)}$
should converge as $x \rightarrow 0$
\cite{heim,ek}.
\item
The isotriplet part of 
$2x g_1$ is significantly greater than the isotriplet
part of the spin-independent structure function $F_2$ 
for $x$ between 0.01 and 0.12.
This result
corresponds to
the parton-model flavour inequality
$(d + {\overline d})^{\downarrow}(x) > (u + {\overline
  u})^{\downarrow}(x)$.
%
\item
We analyse the low $Q^2$ ($\sim 0.5$GeV$^2$) 
data from E-143 \cite{e143b} and SMC
\cite{smc}.
We find that the isosinglet deuteron asymmetry
$A_1^d$ is very small and consistent with zero 
in both experiments
(at $\sqrt{s} \simeq 3.5$ and 16.7GeV)
whereas the E-143 proton data exhibits a clear 
positive proton 
asymmetry $A_1^p$ at $\sqrt{s} \simeq 3.5$GeV.
\end{enumerate}
Finally, we use the low $Q^2$ data to estimate the high-energy 
Regge contribution to the 
Drell-Hearn-Gerasimov sum-rule \cite{dhg}
which is expected to hold at $Q^2=0$.
We find that a combined estimate of nucleon resonance
\cite{ikar}, strangeness \cite{moia} and high-energy 
Regge contributions to the Drell-Hearn-Gerasimov sum-rule
is consistent with the theoretical prediction 
for the fully inclusive sum-rule to within one standard deviation.
We conclude (Section 4) by summarising 
the present status of 
knowledge about high-energy photoabsorption 
and the spin and flavour structure of the nucleon at low $Q^2$.

\section{Sum-rules and spin dependent Regge theory}

\subsection{The asymmetry $A_1$}

The spin dependent photoabsorption cross-sections 
can be expressed as
\begin{equation}
(\sigma_A - \sigma_P) = {4 \pi \alpha^2 \over m {\cal F} }
                        ({g_1 - {Q^2 \over \nu^2} g_2})
\end{equation}
and
\begin{equation}
(\sigma_A + \sigma_P) = {4 \pi \alpha^2 \over m {\cal F} } 
                        F_1.
\end{equation}
Here $g_1$ and $g_2$ are the first and second nucleon spin
dependent structure functions and $F_1$ is the first
spin independent structure function which is measured in
unpolarised scattering; 
${\cal F}$ is the photon flux factor which cancels in the 
asymmetry $A_1$, 
viz
\begin{equation}
A_1 
= {g_1 - {Q^2 \over \nu^2} g_2 \over F_1}
= {g_1 - {2 m x \over \nu} g_2 \over F_1}.
\end{equation}
In this paper we shall consider two limits
\begin{enumerate}
\item
polarised deep inelastic scattering;
\item
high energy photoabsorption at low $Q^2$.
\end{enumerate}
In both these limits
\begin{equation}
A_1 \rightarrow {g_1 \over F_1}.
\end{equation}
There are rigorous sum-rules 
for $(\sigma_A - \sigma_P)$
in deep inelastic scattering ($Q^2 \rightarrow \infty$) and at
$Q^2=0$.

\subsection{The deep inelastic $g_1$ sum-rule}

When $Q^2 \rightarrow \infty$, 
the light-cone operator product expansion relates
the first moment of the structure function $g_1$ 
to the scale-invariant axial charges of the target nucleon
by \cite{kod,larin,altr}
\begin{eqnarray}
\int_0^1 dx \ g_1^p (x,Q^2) &=& 
\Biggl( {1 \over 12} g_A^{(3)} + {1 \over 36} g_A^{(8)} \Biggr) 
\Bigl\{1 + \sum_{\ell\geq 1} c_{{\rm NS} \ell\,}
\bar{g}^{2\ell}(Q)\Bigr\} \nonumber \\
&+& {1 \over 9} g_A^{(0)}|_{\rm inv} 
\Bigl\{1 + \sum_{\ell\geq 1} c_{{\rm S} \ell\,} \bar{g}^{2\ell}(Q)\Bigr\}
\ + \ {\cal O}({1 \over Q^2}).
\end{eqnarray}
The flavour non-singlet $c_{{\rm NS} \ell}$  
and singlet $c_{{\rm S} \ell}$ coefficients
are calculable in 
$\ell$-loop perturbation theory \cite{larin}.
The first moment of $g_1$ is fully constrained 
by low energy 
weak interaction dynamics.
The isotriplet axial charge $g_A^{(3)}$ is measured independently
in neutron beta decays, the flavour octet axial charge
$g_A^{(8)}$ is measured independently in hyperon beta decays, and
the flavour singlet, scale invariant axial charge $g_A^{(0)}|_{\rm inv}$
\cite{mink}
may be measured independently in elastic neutrino proton scattering
\cite{bcft}.

Polarised deep inelastic scattering experiments 
at CERN \cite{expt, smc}, DESY \cite{hermes} and SLAC
\cite{e143a,e143b}
have verified 
the Bjorken sum-rule \cite{bj}
for the isovector part of $g_1$
to within 15\%.
They have also revealed a four standard deviations violation 
of OZI in the flavour singlet axial charge $g_A^0|_{\rm inv}$ 
---
for recent reviews see \cite{altr, cheng}.
The 
small $x$ extrapolation of $g_1$ data 
is presently the largest source of experimental error on 
measurements of the nucleon's axial charges from deep inelastic 
scattering.

\subsection{The Drell-Hearn-Gerasimov sum-rule}

For real photons, $Q^2=0$,
the Drell-Hearn-Gerasimov sum-rule \cite{dhg}
(for reviews see \cite{moia, drec})
relates the difference of 
$\sigma_A$ and 
$\sigma_P$ 
to the square of the nucleon's anomalous magnetic moment
\begin{equation}
({\rm DHG}) \equiv
- {4 \pi^2 \alpha \kappa^2 \over 2 m^2} =
\int_{\nu_{th}}^{\infty} {d \nu \over \nu} (\sigma_A - \sigma_P)(\nu).
\end{equation}
Here $\nu$ is the energy of the incident photon in the target rest
frame,
$m$ is the nucleon mass and $\kappa$ is
the anomalous magnetic moment.

Charge parity imposes a symmetry constraint on dynamical
contributions to the Drell-Hearn-Gerasimov integral.
The cross-sections $\sigma_A$ and $\sigma_P$ are even under charge
parity
($C=+ 1$).
They receive contributions from OZI violating processes where the 
photon couples to the target nucleon via a $C=+1$, colour neutral, 
gluonic intermediate 
state in the $t$ channel, for example two gluon exchange.
The anomalous magnetic moment $\kappa$ which appears on the left 
hand side of the Drell-Hearn-Gerasimov sum-rule has charge parity 
$C= -1$.  
It is measured in the nucleon matrix element of the vector current.
Processes which contribute to $C= +1$ observables 
but not to matrix 
elements of
the conserved $C=-1$ vector current
therefore cancel 
in the logarithmic 
Drell-Hearn-Gerasimov integral for the difference of 
the two spin dependent photoabsorption cross-sections \cite{moia,moib}.

The theoretical predictions for the isoscalar and isovector parts
of the DHG integral 
$({\rm DHG})_{(I=0,1)}^{\rm inclusive}$
are:
\begin{equation}
({\rm DHG})_{I=0}^{\rm inclusive} = -219{\rm \mu b},   \ \ \ \ \ \ 
({\rm DHG})_{I=1}^{\rm inclusive} = + 15{\rm \mu b}.   
\end{equation}
The first direct measurements of $\sigma_A$ and $\sigma_P$ at $Q^2=0$
will soon be available from the ELSA, GRAAL, LEGS and MAMI
facilities 
up to $\sqrt{s_{\gamma p}} \simeq 2.5$GeV.

Multipole analyses \cite{ikar}
of unpolarised pion photoproduction data suggest that 
the isosinglet part of the Drell-Hearn-Gerasimov
sum-rule may be nearly saturated by nucleon resonance
contributions
with estimates ranging between
-225${\rm \mu b}$ and $ \ -222{\rm \mu b}$.
Estimates of the nucleon resonance contribution 
to the isovector part 
to the DHG
integral
range between -65${\rm \mu b}$ and -39${\rm \mu b}$  
-- that is,   different in sign and a factor of 2-4 
bigger than the theoretical prediction 
for the isovector part of the fully inclusive sum-rule.

To understand the ``discrepancy'' between the inclusive
(DHG) sum-rule and the nucleon resonance contributions
to the inclusive sum-rule, it is helpful 
to express the anomalous magnetic moment
$\kappa_p$ as the sum of
its isovector $\kappa_3$ and isoscalar $\kappa_0$ 
parts
($\kappa_p = \kappa_0 + \tau_3 \kappa_3$).
We can then write separate isospin sum-rules:
\begin{eqnarray}
  ({\rm DHG})_{I=0} = ({\rm DHG})_{[33]} + ({\rm DHG})_{[00]}
                    &=& - {2 \pi^2 \alpha \over m^2}
                         (\kappa_3^2 + \kappa_0^2)     \\ \nonumber
  ({\rm DHG})_{I=1} = ({\rm DHG})_{[30]}
                    &=& - {2 \pi^2 \alpha \over m^2}
                         2 \kappa_3 \kappa_0.
\end{eqnarray}
The physical values of the proton and nucleon anomalous
magnetic moments $\kappa_p = 1.79$ and $\kappa_n = -1.91$
correspond to
$\kappa_0 = -0.06$ and $\kappa_3 = +1.85$.
If we take the Drell-Hearn-Gerasimov sum-rule as an exact 
equation and just the pion photoproduction data, 
then we can invert this equation to extract the 
following 
``pion physics'' contributions to the anomalous magnetic 
moment:
$\kappa_0^{(\pi)}$ between +0.16 and +0.26,
and
$\kappa_3^{(\pi)} = +1.86$.

Strange quark dynamics are not expected to play 
an important role in
pion-photoproduction.
The ``pion physics'' contribution $\kappa^{(\pi)}$
to the anomalous magnetic moment
is not expected to receive any large contribution
from the strangeness part,
$-{1 \over 3} G_M^s(0)$, 
of the anomalous magnetic moment \cite{moia}.
(Here $G_M^s(Q^2)$ is the strangeness magnetic
 form-factor.)
The SAMPLE experiment at Bates have recently measured
the $G_M^s(Q^2)$ at
$Q^2=0.1$GeV$^2$ \cite{sample}.
They find
$G_M(0.1 {\rm GeV}^2) = +0.23 \pm 0.44$.
If this measurement is extrapolated as a constant 
to $Q^2=0$,
then the SAMPLE measurement
corresponds to a strangeness
contribution $-0.08 \pm 0.15$ 
to the isoscalar part of the anomalous magnetic moment.
Substituting in Eq.(9), 
we estimate the strangeness contributions
to the isovector and isoscalar
Drell-Hearn-Gerasimov integrals 
as
$+20 \pm 38 \mu$b 
and
$\simeq 1 \mu$b 
respectively.

The positive value of $G_M^s$ measured by SAMPLE
suggests that strangeness helps to fill part of 
the non-resonant contribution to the (DHG) integral.
The high-energy part of $(\sigma_A - \sigma_P)$ 
is expected to behave according 
to spin dependent Regge theory.

\subsection{Spin dependent Regge theory}

At large centre of mass energy squared ($s = 2m \nu + m^2$),
soft Regge theory predicts \cite{heim,ek,peie,clos1,sbpl,kuti}
\begin{equation}
\Biggl( \sigma_A - \sigma_P \Biggr) \sim
N_3 s^{\alpha_{a_1} - 1} + N_0  s^{\alpha_{f_1} - 1}
+ N_g {\ln {s \over \mu^2} \over s}
+ N_{PP} {1 \over \ln^2 {s \over \mu^2} }.
\end{equation}
Here $\alpha_{a_1}$ and $\alpha_{f_1}$
are the intercepts of the isovector
$a_1(1260)$ and isoscalar $f_1(1285)$ and $f_1(1420)$
Regge trajectories.
If we make the usual assumption that the $a_1$ and $f_1$
trajectories are straight lines running approximately
parallel to the
$(\rho, \omega)$ trajectories then 
they can be continued to arbitrary 
real $J$ by
\begin{equation}
\alpha (t) = \alpha_0 + \alpha ' t
\end{equation}
where $\alpha' \simeq +0.86$GeV$^{-2}$.
If we then average over the masses of 
the three $J^{PC} = 1^{++}$
mesons,
we find an average intercept
$\alpha_0 = \alpha(0) = -0.5$.
Alternatively, if we assume a linear $a_1$ 
trajectory running
through the
$a_1(1260)$ and the listed, 
but not well established,
$J^{PC}=3^{++}$ state $a_3(2050)$ \cite{pdg},
then we find a trajectory
with slope
$\alpha'_{a_1} \simeq +0.76$GeV$^{-2}$ 
and intercept
$\alpha_{a_1} = -0.2$.
Both of these two estimates of the $a_1$ intercept
lie within the range
($-0.5 \leq \alpha_{a_1} \leq 0$) quoted in Ref.\cite{ek}. 

The isosinglet part of $(\sigma_A - \sigma_P)$ 
receives a contribution which depends 
on the Lorentz structure of the short range 
exchange potential \cite{clos1}.
If the short range exchange potential which 
generates the pomeron is a scalar, then it 
will not contribute to the large $\sqrt{s}$
limit of $(\sigma_A - \sigma_P)$.
In the Landshoff-Nachtmann approach \cite{pvl1,pvl2},
the soft pomeron is modelled by the exchange of
two non-perturbative gluons
and transforms as a $C=+1$ vector potential
with a correlation length of about 0.1fm.
This vector two non-perturbative gluon exchange 
gives \cite{sbpl} the
$(\ln s) / s$ term in Equ.(2).
The $1 / \ln^2 s$ term represents any 
two-pomeron cut
contribution to $(\sigma_A - \sigma_P)$ \cite{kuti}.

The coefficients $N_3$, $N_0$, $N_g$ and $N_{PP}$ 
in
Equ.(10) are to be determined from experiment.
Each of the possible Regge contributions 
in Equ.(10) yield a convergent Drell-Hearn-Gerasimov integral.
The mass parameter $\mu$ is taken as a typical hadronic
scale (between 0.2 and 1.0GeV).

It is an open question how far we can increase $Q^2$ and still 
trust soft Regge theory to provide an accurate description of
$(\sigma_A - \sigma_P)$.
The HERA measurements \cite{H1, Zeus} of
$(\sigma_A + \sigma_P)$ at low $Q^2$ 
suggest that soft Regge theory provides a good description of
the large $\sqrt{s}$ behaviour
of the total photoabsorption cross-section up to $Q^2 \simeq 
0.5$GeV$^2$.
The shape of $g_1$ at small $x$ is $Q^2$ dependent, 
being driven by DGLAP evolution 
\cite{bfr,abfr,gehr,reya} 
and 
by
the resummation of $\alpha_s^{l+1} \ln^{2l} x$
radiative corrections
\cite{kirs,bart,blum,bade,koda}
, which are expected to play an important
role at very small $x$
(below $x_{\rm min} \simeq 0.005$, the present smallest
 $x$ data from SMC \cite{smc}).

\section {Results from experiment}

We now examine present experimental data with a view 
to extracting information about 
the large $\sqrt{s}$ behaviour of $\sigma_A$ and $\sigma_P$.
We begin our discussion with polarised deep inelastic data 
(Section 3.1)
and then consider the low $Q^2$ measurements from E-143 and 
SMC (Section 3.2).

\subsection{$g_1$ and $F_2$}

Polarised deep inelastic data
from
CERN, DESY and SLAC
reveals
a strong isotriplet term in $g_1$
at small $x$
($x < 0.15 $).
The SLAC E-143 proton \cite{e143a}, 
E-154 neutron \cite{e154}
and 
preliminary E-155 proton \cite{e155} data 
have the smallest experimental error within
the particular kinematic region of the various experiments.
We combine the E-154 neutron data with the proton data
from E-143 and the preliminary E-155 
$x = (0.016, 0.024)$ data points
\footnote{
 Our data set consists of E-143 data which was evolved 
 to $Q^2=$ 3GeV$^2$ 
 and
 E-154 and E-155 data which was evolved to $Q^2=$ 5GeV$^2$
 by assuming $A_1$ is independent of $Q^2$.
 DGLAP evolution is expected to induce some $Q^2$ 
 dependence in the deep inelastic $A_1$. However,
 the $Q^2$ independent hypothesis is consistent with
 the present deep inelastic data on $A_1$.
 Following Soffer and Teryaev \cite{soff} we combine 
 the E-143 and E-154 and E-155 data as if
 they were taken at the same $Q^2$.  The theoretical
 error induced by this procedure is of the order of
 10\%; it is small compared to the present experimental
 error on the data.}.
Assuming that $g_1^{(p-n)}=(g_1^p - g_1^n)$ 
has a power behaviour, 
$x^{\lambda}$,
at small $x$
we find a best fit to the isotriplet part of $g_1$ :
\begin{equation}
g_1^{(p-n)}
 \sim (0.13) x^{-0.49} \ \ \ \ 
{\rm at}
\ \ \ \ 
(0.01 < x < 0.123)
\end{equation}
with $\chi^2=2.19$ for 6 degrees of freedom -- see Fig.1.
A similar fit for $g_1^{(p-n)}$
was obtain by Soffer and Teryaev \cite{soff}.
The isotriplet part of $g_1$
behaves as
$x^{-{1 \over 2}}$ 
at small $x$ in the SLAC data corresponding 
to an effective Regge intercept
$\alpha_{a_1}(Q^2) \simeq +{1 \over 2}$ at
the relatively low deep inelastic $Q^2 \simeq 3-5$GeV$^2$.
This compares with the soft Regge
prediction that $\alpha_{a_1}(0)$ 
is between $-{1\over 2}$ and 0.
If we extrapolate Equ.(12) to $x=0$, then the small $x$ 
(less than 0.12) part of $g_1^{(p-n)}$ contributes 50\% 
of the Bjorken sum-rule. 

In Fig. 1 we also show the SLAC data on 
$g_1^{(p+n)}= (g_1^p + g_1^n)$.
We make a fit to this data by assuming a linear 
combination of a power term $x^{\lambda}$ and 
the functional form $(2\ln{1 \over x} -1)$
which was derived \cite{sbpl}
in the two non-perturbative gluon exchange model \cite{pvl2}.
We find:
\begin{equation}
g_1^{(p+n)} \sim - (0.23) x^{-0.56} + (0.28) (2 \ln {1 \over x} - 1)
 \ \ \ \ {\rm at}
\ \ \ \
(0.01 < x < 0.123)
\end{equation}
with $\chi^2=2.95$ for 6 degrees of freedom
\footnote{
Lower $x$ data is available from SMC \cite{smc} with
larger experimental errors.
We note the isotriplet data points
$(x, g_1^{(p-n)})$:
$(0.005, 1.1 \pm 1.3)$, $(0.008, 2.4 \pm 0.8)$ and
$(0.014, 1.5 \pm 0.5)$
and isosinglet points
$(x, g_1^{(p+n)})$:
$(0.005, 0.0 \pm 1.3)$, $(0.008, -0.5 \pm 0.8)$ and
$(0.014, -0.7 \pm 0.5)$
evolved \cite{gspin} to a common $Q^2 = 5$GeV$^2$.}
.
If we keep only a power term in the fit to the SLAC
$g_1^{(p+n)}$, 
then we find $g_1^{(p+n)} \sim (0.35) x^{+0.36}$ 
with $\chi^2=7.1$ for 6 degrees of freedom.
The power exponent $\lambda$ is negative in the fits (12) 
and (13)
in contrast to the soft Regge prediction $\lambda \geq 0$.
If we extrapolate Equ.(13) to $x=0$ 
and evaluate the small $x$ 
($\leq 0.12$) contribution to the first 
moment of 
$g_1^{(p+n)}$
then we find -0.21 from the power term and 
+0.18 from the $(2\ln{1 \over x} -1)$ term.
This result compares with the 
parton model 
analysis \cite{abfr}
which suggests that the polarised gluon distribution makes
a negative contribution to $g_1$
at small $x$.

Whilst Eq.(13) provides a good fit to the 
data,
it should be treated with care.
The fitted
$g_1^{(p+n)}$ 
is the sum of two terms with opposite sign,
each of which is five times bigger in 
magnitude than the measured $g_1^{(p+n)}$.
As noted in \cite{strat,leader} 
the decomposition of $g_1^{(p+n)}$ 
into the sum of a quark term and a
gluonic 
term is not well constrained at the present time.
Our study supports this observation.
The data in Fig.1 suggests either that $g_1^{(p+n)}$ 
changes sign at 
$x \sim 0.03$
or that it is very close to zero for $x < 0.03$.
Let us suppose that
$g_1^{(p+n)}$ does change sign around $x \sim 0.03$.
The two functional forms that 
we use in the fit (13)
are both positive definite at $x$ below 0.12.
The power term 
has a negative coefficient in Equ.(13) 
because it grows 
faster 
than the logarithmic $(2\ln{1 \over x}-1)$ 
term 
at small $x$.
If the $(2 \ln{1 \over x} -1)$ term were
replaced in the fit by a different 
functional form
which grows faster 
with decreasing $x$
than the 
power contribution $x^{-{1 \over 2}}$,
then
the signs of the coefficients of 
the power and gluonic exchange terms would be reversed.
A direct measurement of the sign and shape of
the polarised gluon distribution 
will come from experimental studies of 
charm \cite{compass} 
and 
di-jet production \cite{HERA, radel} 
in polarised deep inelastic scattering.

To further understand the spin and flavour structure 
at small $x$ it is
helpful to compare
the isotriplet parts of $g_1$ and $F_2$.
In the parton model
\begin{equation}
2x (g_1^p - g_1^n) = 
{1 \over 3} x 
\Biggl[ (u + {\overline u})^{\uparrow} - (u + {\overline u})^{\downarrow} 
      - (d + {\overline d})^{\uparrow} + (d + {\overline
         d})^{\downarrow}
 \Biggr] \otimes \Delta C_{NS}
\end{equation}
and
\begin{equation}
(F_2^p - F_2^n) 
= 
{1 \over 3} x 
\Biggl[ (u + {\overline u})^{\uparrow} + (u + {\overline u})^{\downarrow} 
      - (d + {\overline d})^{\uparrow} - (d + {\overline
         d})^{\downarrow}
 \Biggr] \otimes C_{NS}.
\end{equation}
Here $u$ and $d$ denote the up and down flavoured quark 
distributions
polarized parallel $({\uparrow})$ and antiparallel $({\downarrow})$
to the target proton
\footnote
 {Equs.(14) and (15) assume that charge symmetry is exact in
  relating the $u$ and $d$ flavoured distributions in the
  proton to the $d$ and $u$ distributions in the neutron.
  Possible charge symmetry violations of up to 3-5\% between
  the unpolarised valence distributions 
  $d^p_V(x)$ and $u^n_V(x)$ are discussed in \cite{isospin}.};
$\Delta C_{NS}$ and
$C_{NS}$ denote
the spin-dependent and spin-independent 
perturbative QCD coefficients respectively.
These coefficients are related 
(in the ${\overline{\rm MS}}$ scheme)
by \cite{ratc}
\begin{equation}
\Delta C_{NS} (x) = C_{NS}(x) - {\alpha_s \over 2 \pi}
{4 \over 3}(1 + x).
\end{equation}

In Fig. 2
we compare the isotriplet part of the SLAC $2xg_1$ data 
with the NMC
measurement \cite{nmc}
of the isotriplet part of $F_2$ at 4GeV$^2$.
The NMC parametrised their small $x$ data using the fit:
\begin{equation}
(F_2^p - F_2^n) \sim (0.20 \pm 0.03) x^{0.59 \pm 0.06} \ \ \ \ 
{\rm at} \ \ \ \ 
(0.004 < x < 0.15)
\end{equation}
for $Q^2=4$GeV$^2$ .
Clearly, the isotriplet part of $2x g_1$ is significantly
greater than the isotriplet part of $F_2$ 
\begin{equation}
2x (g_1^p - g_1^n) > (F_2^p - F_2^n)
\end{equation}
indicating the flavour inequality
\begin{equation}
(d + {\overline d})^{\downarrow} (x) 
> 
(u + {\overline u})^{\downarrow} (x) \ \ \ \ 
{\rm at} \ \ \ \ 
(0.01 < x < 0.12).
\end{equation}
This inequality includes both valence and sea contributions.
It holds both at leading order and, via Equ.(16), 
also at next-to-leading order.
Since the coefficient $C_{NS}$ is greater than 
 $\Delta C_{NS}$ at next-to-leading order, 
 Equ.(16), it follows that the the parton-model 
 inequality (19) is more pronounced
 at next-to-leading order than at leading order.
(The $- {\alpha_s \over 2 \pi} {4 \over 3}(1+x)$
 term in Equ.(16) acts to lower $2x g_1^{(p-n)}$
 towards $F_2^{(p-n)}$ in Fig.2.)
The inequality (18) persists in the data to $x \simeq 0.4$.

One also finds that
the measured \cite{nmc} Gottfried integral \cite{gott}
\begin{equation}
\int_0^1 dx \Biggl( {F_2^p - F_2^n \over x} \Biggr)
=  0.235 \pm 0.026
\end{equation}
at $Q^2=4$GeV$^2$ is five standard deviations 
below the theoretical value of 
the Bjorken sum-rule \cite{bj, larin}
\begin{eqnarray}
2 \int_0^1 dx \Biggl( g_1^p - g_1^n \Biggr)
&=&
\frac{g_A^{(3)}}{3} \left[1 - \frac{\alpha_s}{\pi} - 3.58 \left(
\frac{\alpha_s}{\pi} \right)^2 - 20.21 
\left(\frac{\alpha_s}{\pi} \right)^3 \right] \\
&=&  0.370 \pm 0.008
\end{eqnarray}
at $Q^2=4$GeV$^2$.

\subsection{$A_1$ at low $Q^2$}

The E-143 \cite{e143b} and SMC \cite{smc}
experiments have measured $A_1$ 
for both proton and deuteron targets
over a wide
range of $Q^2$, 
including between 0.25 GeV$^2$ and 0.80 GeV$^2$.
We list this low $Q^2$ data in Table 1.

The low $Q^2$ data has the following general 
features.
First,
the isoscalar deuteron asymmetry $A_1^d$
is very small and consistent with zero 
in both the E-143 and SMC low $Q^2$ bins.
Second,
there is a clear positive proton asymmetry
in the
E-143 data,
signalling a strong isotriplet term
in $(\sigma_A - \sigma_P)$ at $s \simeq 12$GeV$^2$.
The SMC $A_1^p$ data is less clear:
combining the SMC low $Q^2$ $A_1^p$ data yields a
positive value for $A_1^p$.
However, the majority of these SMC points 
are consistent with zero and more precise 
data are
needed to resolve this value.
\begin{table}
\begin{center}
\caption{Small $Q^2$ data from E-143 and SMC } 
\begin{tabular} {ccccc}
\\
\hline\hline
\\
$x$ & $Q^2$ & $s$ & $A_1^p$ & $A_1^d$ \\
\\
\hline
\\
SLAC E-143 & & & & \\
\hline
0.035  &  0.32  &   9.7  &  $ 0.053 \pm 0.030 $  & $ -0.020 \pm 0.032 $ \\
0.035  &  0.65  &  18.8  &  $ 0.069 \pm 0.018 $  & $ +0.039 \pm 0.046 $ \\
0.050  &  0.37  &   7.9  &  $ 0.110 \pm 0.033 $  & $ +0.004 \pm 0.034 $ \\
0.050  &  0.79  &  15.9  &  $ 0.117 \pm 0.019 $  & $ +0.023 \pm 0.034 $ \\
0.080  &  0.42  &   5.7  &  $ 0.095 \pm 0.037 $  & $ +0.031 \pm 0.040 $ \\    
0.080  &  0.71  &   9.0  &  $ 0.129 \pm 0.038 $  & $ -0.010 \pm 0.043 $ \\  
0.125  &  0.47  &   4.2  &  $ 0.110 \pm 0.048 $  & $ +0.022 \pm 0.057 $ \\
\hline\hline
CERN SMC & & & & \\
\hline
0.0009 &  0.25  &   278  &  $ 0.001 \pm 0.069 $  & $ -0.058 \pm 0.055 $ \\
0.0011 &  0.30  &   273  &  $ 0.016 \pm 0.085 $  & $ +0.004 \pm 0.067 $ \\
0.0011 &  0.34  &   309  &  $ 0.196 \pm 0.111 $  & $ +0.056 \pm 0.084 $ \\
0.0014 &  0.38  &   272  &  $ 0.139 \pm 0.044 $  & $ -0.045 \pm 0.041
$ \\
0.0017 &  0.46  &   271  &  $ 0.076 \pm 0.053 $  & $ -0.088 \pm 0.050
$ \\
0.0019 &  0.55  &   290  &  $ 0.037 \pm 0.057 $  & $ +0.013 \pm 0.055
$ \\
0.0023 &  0.58  &   252  &  $ 0.020 \pm 0.040 $  & $ +0.114 \pm 0.042
$ \\
0.0025 &  0.70  &   280  &  $ 0.025 \pm 0.044 $  & $ -0.099 \pm 0.046
$ \\
\hline\hline
\end{tabular}
\end{center}
\end{table}

Due to the wide separation in $s$ range measured in E143 and SMC, 
we combine the low $Q^2$ data to obtain one point corresponding 
to each experiment.
This is shown in Table 2.
We make two cuts:
\begin{enumerate}
\item
keeping $\sqrt{s} \geq 2.5$GeV to ensure that our data set is well
beyond the resonance region and 
including all such data that 
the mean $Q^2$ is kept below 0.5GeV$^2$ for each experiment.
(In practice, this amounts to a common $Q^2$ cut of
  0.7GeV$^2$ and 
  yields a mean $Q^2=0.45$GeV$^2$ for each experiment.)
\item
including all data at $\sqrt{s} \geq 2$GeV and
$Q^2 \leq 0.8$GeV$^2$.
\end{enumerate}
The HERMES experiment hope to measure the low $Q^2$
$A_1^p$ at $\sqrt{s} \simeq 7$GeV over the same
range of $Q^2$ and with similar accuracy to the
SLAC experiments \cite{brull}.
\begin{table}
\begin{center}
\caption{$A_1$ at large $s$ and low $Q^2$}
\begin{tabular} {ccccc}
\\
\hline\hline
\\
Cuts & $\langle Q^2 \rangle$ & $s$ & $A_1^p$ & $A_1^d$ \\
\\
\hline
\\
(a) \
$\langle Q^2 \rangle \leq 0.5$GeV$^2$, $s \geq 7$GeV$^2$ & & & & \\
\hline
  &  0.45  &  12  &  $ 0.077 \pm 0.016 $  & $ +0.008 \pm 0.022 $ \\
  &  0.45  &  278 &  $ 0.064 \pm 0.024 $  & $ -0.013 \pm 0.020 $ \\
\hline\hline
\\
(b) \
$Q^2 \leq 0.8$GeV$^2$, $s \geq 4$GeV$^2$ & & & & \\
\hline
  &  0.53  &  10  &  $ 0.098 \pm 0.013 $  & $ +0.016 \pm 0.016 $ \\
  &  0.45  &  278 &  $ 0.064 \pm 0.024 $  & $ -0.013 \pm 0.020 $ \\
\hline\hline
\end{tabular}
\end{center}
\end{table}

In what follows, we work with Cut (a).
This choice of cut is a compromise between keeping
$Q^2$ as low as possible and including the maximum
amount of data.
The choice $Q^2_{\rm max} \simeq 0.5$GeV$^2$
is motivated by the HERA data \cite{H1,Zeus}
on $(\sigma_A + \sigma_P)$ 
which rises with increasing $\sqrt{s}$
according to soft Regge theory up to 
$Q^2 \simeq 0.5$GeV$^2$.
At larger $Q^2$ the data exhibits evidence of
$Q^2$ dependence in the
effective 
Regge intercepts 
for high-energy, virtual photoabsorption.

To estimate the spin asymmetry at $Q^2=0$ we shall 
assume that 
the large $\sqrt{s}$ $A_1$ is approximately
independent of $Q^2$
between $Q^2=0$ and $Q^2 \simeq$0.5 GeV$^2$.
Since the E-143 
data has the lowest experimental error and shows a
clear positive signal in $A_1^p$
at low $Q^2$
we choose to normalise to E-143.
For the total photoproduction cross-section we take
\begin{equation}
(\sigma_A + \sigma_P) = 67.7 s^{+0.0808} + 129 s^{-0.4545}
\label{sigtot}
\end{equation}
(in units of $\mu$b),
which is known to provide a good Regge fit for $\sqrt{s}$
between 2.5GeV and 250GeV \cite{pvl1}.
(Here,
the $s^{+0.0808}$ contribution is associated with pomeron
exchange
and the $s^{-0.4545}$ contribution is associated with the
isoscalar $\omega$ and isovector $\rho$ trajectories.)
Multiplying $A_1^p$ by the value of $(\sigma_A + \sigma_P)$
at $\sqrt{s}=3.5$GeV,
we estimate
\begin{equation}
(\sigma_A - \sigma_P) \simeq +10 \mu{\rm b} \ \ \ \
{\rm at} \ \ \ \ (Q^2=0, \sqrt{s} = 3.5 {\rm GeV}).
\end{equation}

The small isoscalar deuteron asymmetry $A_1^d$ indicates that
the isoscalar contribution to $A_1^p$ in the E-143 data is unlikely
to be more than 30\%.
In Fig. 3 we show the asymmetry $A_1^p$ as a function of $\sqrt{s}$
between 2.5 and 250 GeV
for the
four different would-be Regge behaviours for
$(\sigma_A - \sigma_P)$:
that the high energy behaviour of $(\sigma_A-\sigma_P)$ is given
\begin{enumerate}
\item
entirely by the $(a_1, f_1)$ terms in Equ.(2) with Regge
    intercept either (1)
    $-{1 \over 2}$ (conventional) or
    (2)  $+{1 \over 2}$
    (motivated by the observed small $x$ behaviour of $g_1^{(p-n)}$),
\item
by taking 2/3 isovector (conventional) $a_1$ and
1/3 two non-perturbative
gluon exchange contributions at $\sqrt{s} = 3.5$GeV,
\item
by taking 2/3 isovector (conventional) $a_1$ and 1/3
pomeron-pomeron cut
contributions at $\sqrt{s} = 3.5$GeV.
\end{enumerate}
(In Fig.3 we take the mass parameter in the Regge fit,
 Equ.(10), as $\mu^2 = 0.5$GeV$^2$.)

The combined E-143 and SMC data 
are consistent with $A_1^p$ 
being constant in $\sqrt{s}$.
They are consistent at the level of two standard deviations
with 
$(\sigma_A - \sigma_P) \sim s^{-{1 \over 2}}$.
However, given the large experimental error on the SMC data
it is not possible to draw any meaningful conclusion at the 
present stage.

The small isosinglet asymmetry $A_1^d$ is consistent 
with the symmetry constraints from charge parity 
that we discussed in Sect.2.3.
Furry's theorem tells us that two (non-perturbative)
gluon exchange does not contribute to the anomalous
magnetic moment.
It follows that
this exchange yields zero contribution to the (DHG) 
integral. 
The contribution of such processes 
to
$(\sigma_A - \sigma_P)$
is either exactly zero 
or it changes sign (at least once) 
so that they yield a vanishing contribution to (DHG).
The Regge contribution $(\ln s) / s$ 
is positive definite at large $\sqrt{s}$.
The sign of this gluonic exchange 
contribution to $(\sigma_A - \sigma_P)$
is unknown at energies below the Regge region.
The small value of $A_1^d$ 
in the E-143 and SMC low $Q^2$ data 
is consistent with a
vanishing $C=+1$ gluonic exchange
contribution to  $(\sigma_A - \sigma_P)$ at $Q^2=0$.

To estimate the high-energy 
Regge contribution 
to the Drell-Hearn-Gerasimov integral
we fit a Regge form
$(\sigma_A - \sigma_P) \sim s^{\alpha -1}$
through the E-143 value, Equ.(24),
and allow $\alpha$ to vary
between
$\pm {1 \over 2}$.
This leads to an estimate
of
the Regge contribution
to (DHG)
of $+25 \pm 10 \mu$b
from $\sqrt{s} \geq 2.5$GeV taking into account the error on
$A_1^p$.
If we allow a maximum 30\% contribution from the two-pomeron
cut in the E-143 $(\sigma_A - \sigma_P)$
we find a contribution to (DHG) which is at the upper limit
of this range.
(We note, however, that there is no evidence for any
  two-pomeron contribution in present deep inelastic data
  \cite{smc}.)

\section{Conclusions}

We have examined the present data on polarised
photoabsorption 
for $Q^2$ between 0.25 and 5GeV$^2$.

We find two interesting results in the isotriplet
channel.
First,
soft Regge theory predicts that the isotriplet
$(\sigma_A - \sigma_P) \sim s^{\alpha_{a_1}-1}$
at large $s$
and that
$g_1^{(p-n)} \sim x^{- \alpha_{a_1}}$ at small
$x$,
where $(-{1\over 2} \leq \alpha_{a_1} \leq 0)$.
Polarised deep inelastic data 
suggests that
$g_1^{(p-n)} \sim x^{-{1 \over 2}}$ 
for $x$
between 0.01 and 0.12 and $Q^2 \simeq 4$GeV$^2$.
That is, 
$\alpha_{a_1}(Q^2)$ at $Q^2 \simeq 4$GeV$^2$
has the opposite sign to the soft Regge prediction.
It is a challenge for future polarisation
experiments to map the $Q^2$ dependence of
the effective $a_1$ intercept: 
either to observe it change sign at a 
particular value of $Q^2$ or, if does not,
to observe a
Regge trajectory that is either nearly
flat or non-linear at $t \rightarrow 0^+$.
Second, we find the novel result that the 
isotriplet part of the polarised $2x g_1$ 
is significantly greater than the 
isotriplet
part of the unpolarised $F_2$ at small $x$.

In the isosinglet channel, the sign of the 
gluonic exchange contribution to the first 
moment of $g_1$ is not well constrained at
the present time. 
The sign that we extract from the fit 
to $g_1^{(p+n)}$
is sensitive to the functional form 
that is assumed for this term in the fit.

At the present time, there is no data on
$(\sigma_A-\sigma_P)$
at $Q^2=0$.
Experiments with real photons are planned
or underway at the
ELSA, GRAAL, LEGS and MAMI facilities 
to measure the
$(\sigma_A - \sigma_P)$
up to $\sqrt{s_{\gamma p}} \leq 2.5$GeV.
They will make a direct measurement of
nucleon resonance and strangeness production
contributions to the Drell-Hearn-Gerasimov
sum-rule.
One future possibility to measure high energy
$\gamma p$ collisions is polarised HERA 
\cite{HERA}.
If realised, this facility could  
measure $A_1^p$
to an accuracy of 0.0003 at $\sqrt{s}$ 
between 50 and 250 GeV
assuming an integrated luminosity
${\cal L}\simeq 500$pb$^{-1}$.
Details of the detector acceptances and 
the expected asymmetries are given in \cite{us1}.

Whilst we wait for direct measurements of
$(\sigma_A - \sigma_P)$,
the Drell-Hearn-Gerasimov sum-rule allows
us to deduce a consistent picture of the
spin structure of the nucleon at $Q^2=0$.

\begin{enumerate}
\item
Multipole analyses of (unpolarised) pion photoproduction data 
suggest that the isosinglet part of the DHG sum-rule 
(-219 $\mu$b) is nearly fully saturated by nucleon resonance 
contributions (to within a few percent), 
whereas resonance contributions to the isotriplet part of the 
sum-rule are a factor of 2-4 times bigger and have 
the opposite sign to the theoretical prediction for the
isotriplet part of the fully inclusive sum-rule ($+15 \mu$b).
\item
Pion photoproduction data does not include any net
strangeness production in the final state. 
Strangeness contributions to (DHG) 
can be quantified through the strangeness
magnetic moment $G_M^s(0)$
which is measured in parity violating elastic
${\vec e}p$ scattering \cite{moia}.
If one assumes that $G_M^s(Q^2)$ is independent of $Q^2$
between zero and 0.1GeV$^2$, then the recent SAMPLE
measurement of
$G_M^s(0.1{\rm GeV}^2) \simeq 0.23 \pm 0.44$ corresponds
to a strangeness contribution
$+20 \pm 38 \mu$b in the isovector (DHG) integral 
and $\simeq 1 \mu$b in the isoscalar (DHG) integral.
\item
Low $Q^2$ measurements of the spin asymmetry
$A_1$ 
reveal a small isosinglet deuteron asymmetry
(consistent with zero)
and a strong isotriplet term in 
$(\sigma_A - \sigma_P)$
at $\sqrt{s} \simeq 3.5$GeV and 
$Q^2$ between 0.25GeV$^2$ and 0.5GeV$^2$.
If we assume that $A_1$ is independent of
$Q^2$ between $Q^2=0$ and $\simeq 0.5$GeV$^2$,
then we can 
estimate 
the Regge contribution to the 
Drell-Hearn-Gerasimov integral 
as
$+25 \pm 10 \mu$b.
This contribution is predominantly isotriplet.
\end{enumerate}
Both the strangeness and the Regge 
contributions appear to reconcile part of the
``discrepancy'' between the nucleon resonance
contribution to the isovector part of the DHG
integral and the theoretical prediction 
for the isovector part of the fully inclusive
sum-rule.
They appear to yield only a very small 
contribution to the isoscalar part of (DHG)
which multipole analyses suggest is nearly
fully saturated by nucleon resonance contributions.
Combining the above estimates of 
nucleon resonance, strangeness and 
Regge contributions,
which are extracted
from three independent experiments,
the net contribution to each of the
isovector and isoscalar parts of the 
Drell-Hearn-Gerasimov sum-rule is consistent 
with the theoretical 
prediction for the fully inclusive (isospin 
dependent) sum-rules to within one standard
deviation.
Clearly, each of these contributions should
be measured directly in a real photon beam
DHG experiment.
However, even at this preliminary stage,
the available experimental evidence is
consistent with the validity of the DHG
sum-rule.

High-energy polarised $\gamma p$ collisions
continue to offer up many unexpected 
surprises.
Experiments with polarised real photons
are just beginning; 
the $Q^2$ dependence of the spin dependent
part of the high-energy
photoabsorption cross-section, 
through the study of
$\alpha_{a_1}(Q^2)$ and $\alpha_{f_1}(Q^2)$,
would open up a new window on the spin
structure of the nucleon.


\vspace{2.0cm}
{\large \bf Acknowledgements: \\}
\vspace{3ex}

It is a pleasure to thank 
M. Gl{\"u}ck, T. Goldman, P.V. Landshoff, 
E. Reya and A.W. Thomas for theoretical discussions and
A. Br\"ull,  A. De Roeck and G. R{\"a}del for 
discussions about present and future experiments.
Support from the Alexander von Humboldt Foundation (SDB) 
and the U.S. Department of Energy (MMB) is gratefully
acknowledged.

\pagebreak

\newpage
\begin{figure}
\centerline{}
\begin{center}
\input{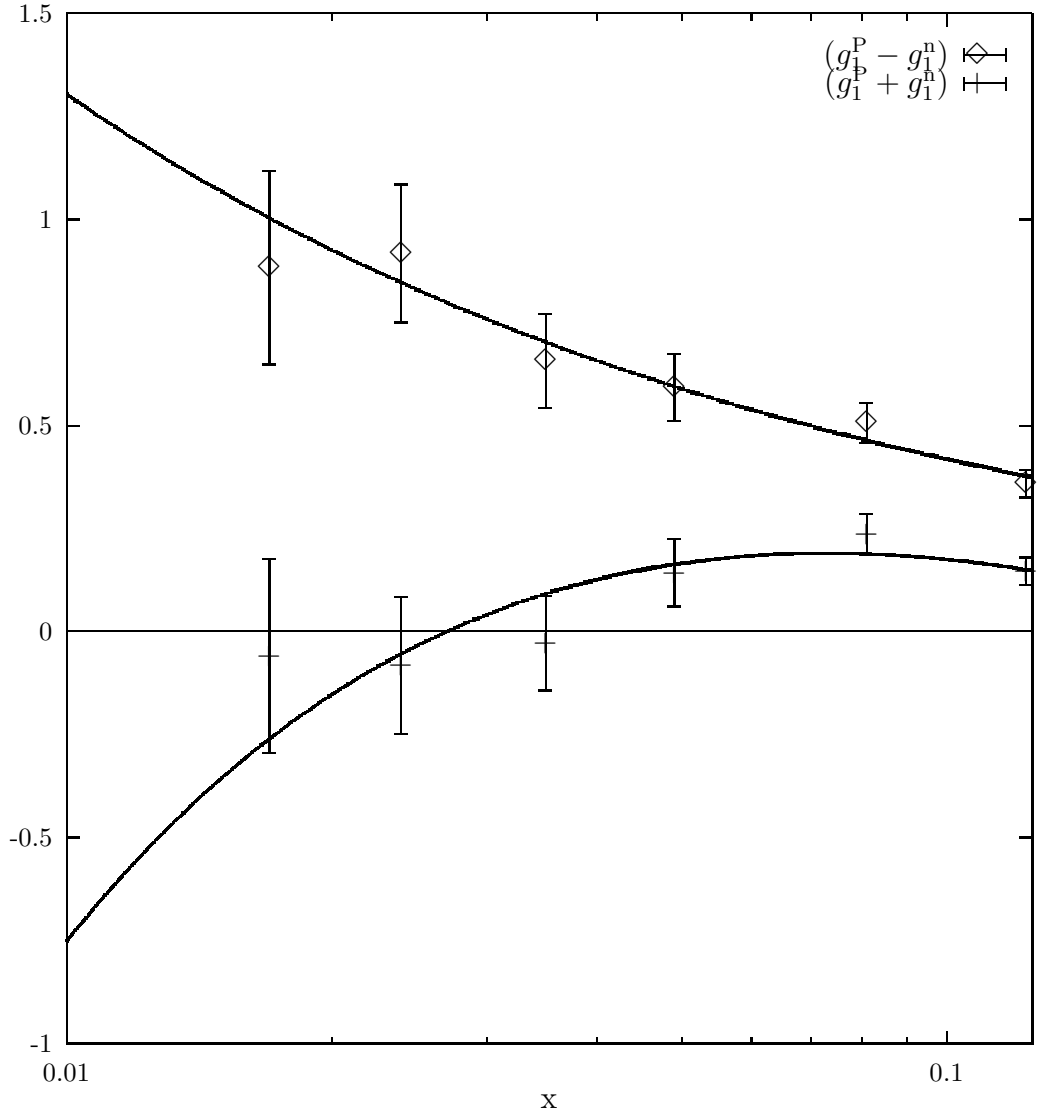}
\caption{
The SLAC data on $g_1$.
The upper curve shows the fit (12) to the isotriplet
$g_1^{(p-n)}(x)$.
The lower curve shows the fit (13) to the isosinglet
$g_1^{(p+n)}(x)$
at $Q^2 \simeq 4$GeV$^2$. 
} 
\end{center}
\end{figure}

\begin{figure}
\centerline{}
\begin{center}
\input{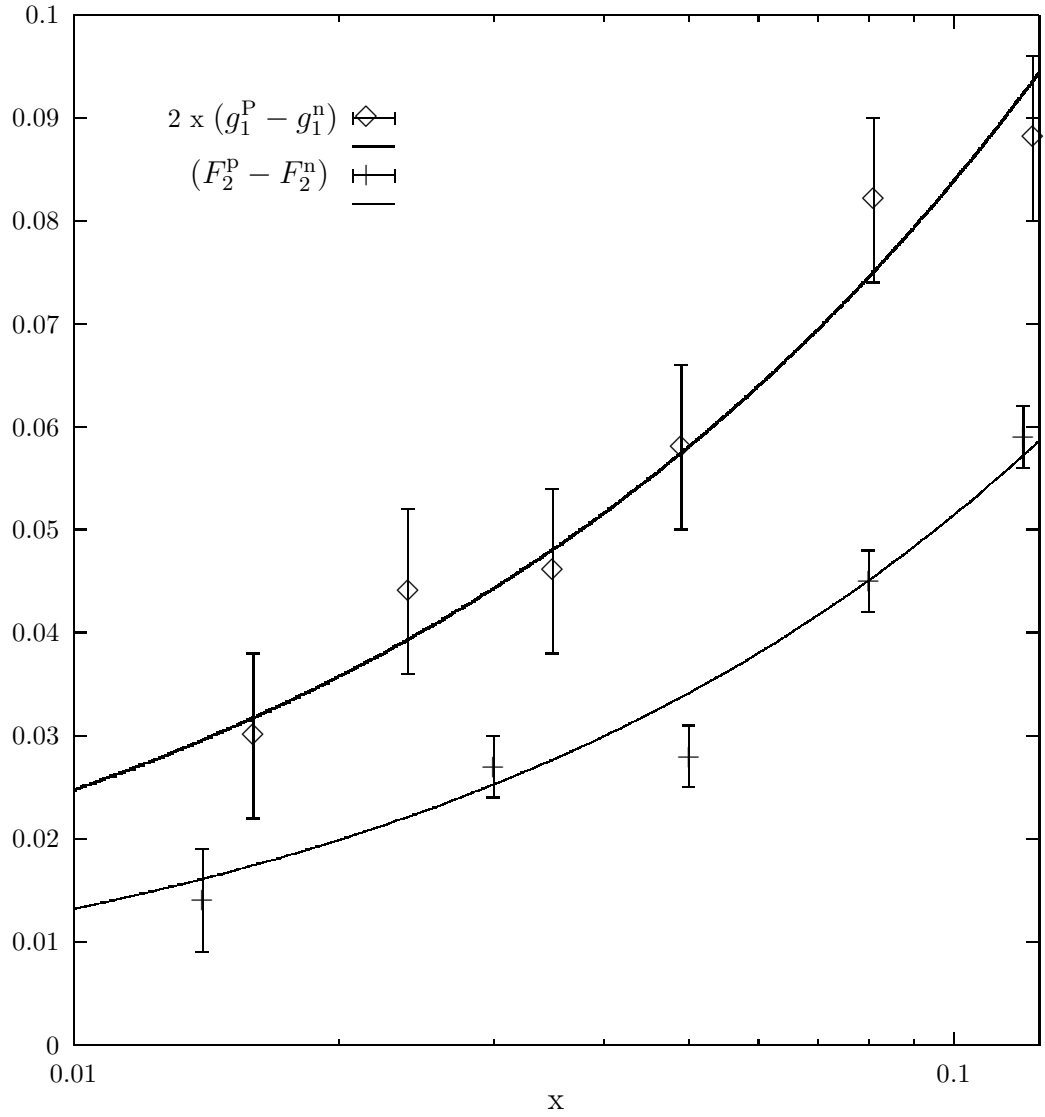}
\caption{
Comparison of the isotriplet parts of the 
polarised 
$2xg_1$ (SLAC) and unpolarised $F_2$ (NMC) 
at $Q^2 \simeq 4$GeV$^2$. 
}
\end{center}
\end{figure}

\begin{figure}
\centerline{}
\begin{center}
\input{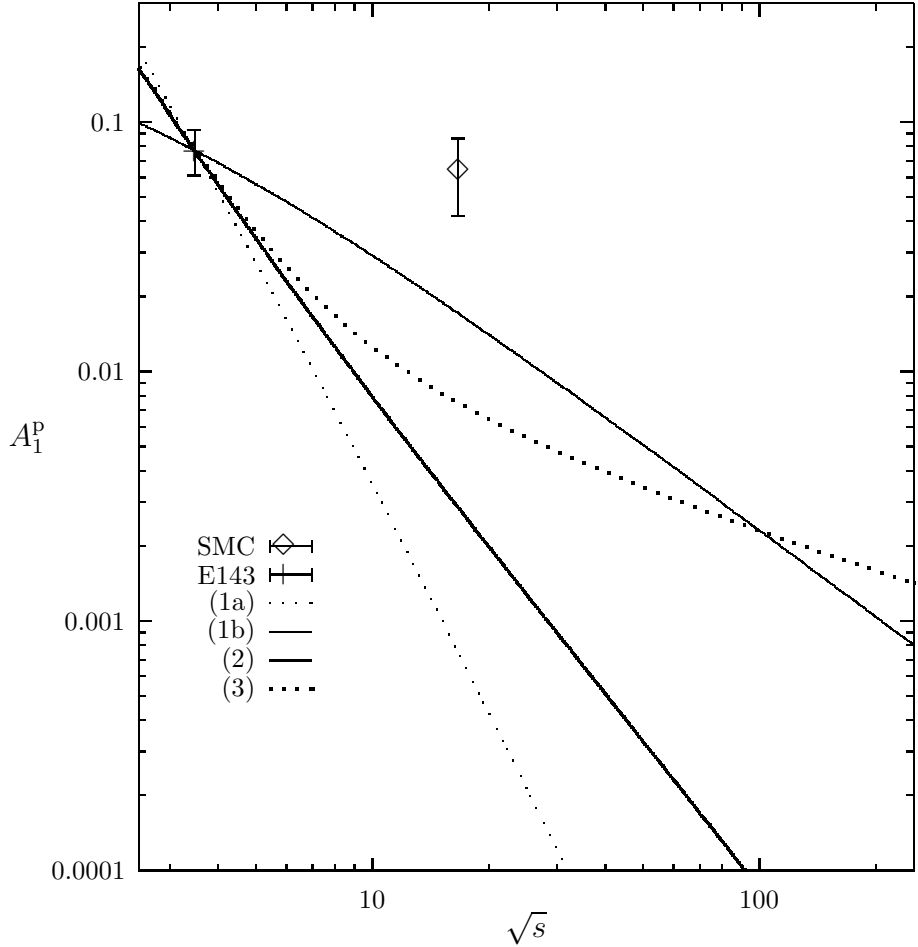}
\caption{
The real photon asymmetry $A_1^p$ as a function of $\sqrt{s}$
for different Regge behaviours for
$(\sigma_A - \sigma_P)$:
 given
entirely by (1a) the $(a_1, f_1)$ terms in Equ.(2) with Regge
    intercept either
    $-{1 \over 2}$ (conventional) or
    (1b)  $+{1 \over 2}$;
(2) by  2/3 isovector (conventional) $a_1$ and
1/3 two non-perturbative
gluon exchange contributions at $\sqrt{s} = 3.5$GeV;
(3)
by 2/3 isovector (conventional) $a_1$ and 1/3
pomeron-pomeron cut
contributions at $\sqrt{s} = 3.5$GeV.  }
\end{center}
\end{figure}

\end{document}